\pdfoutput=1

\documentclass[11pt]{article}

\usepackage[]{acl}
\usepackage{times}
\usepackage{latexsym}
\usepackage[T1]{fontenc}
\usepackage[utf8]{inputenc}
\usepackage{microtype}
\usepackage{inconsolata}
\usepackage{booktabs}
\usepackage{xspace}
\usepackage{amsmath}
\usepackage{amssymb}
\usepackage{mathtools}
\usepackage{tcolorbox}
\usepackage{pgfplots}
\usepackage{caption}
\usepackage{subcaption}
\usepackage[hang,flushmargin]{footmisc}
\usepackage{multirow}
\usepackage{enumitem}

\newcommand{\LRL}[0]{LRL\xspace}

\newcommand{\rv}{Rank\-Vicuna\xspace}
\newcommand{\rg}{Rank\-GPT\xspace}

\newcommand{\rgthreefive}{Rank\-GPT\textsubscript{3.5}\xspace}
\newcommand{\rgfour}{Rank\-GPT\textsubscript{4}\xspace}

\newcommand{\gptthreefive}{GPT\textsubscript{3.5}\xspace}
\newcommand{\gptfour}{GPT\textsubscript{4}\xspace}

\newcommand{\ignore}[1]{}

\begin{document}

\title{\rv: Zero-Shot Listwise Document Reranking\\ with Open-Source Large Language Models}

\author{Ronak Pradeep\thanks{~~~Equal contribution.}$^{*}$, Sahel Sharifymoghaddam$^{*}$, Jimmy Lin \\[1ex]
David R. Cheriton School of Computer Science,\\
University of Waterloo, Canada \\[1ex]
\texttt{\{rpradeep, sahel.sharifymoghaddam, jimmylin\}@uwaterloo.ca}}

\maketitle

\begin{abstract}
Researchers have successfully applied large language models (LLMs) such as ChatGPT to reranking in an information retrieval context, but to date, such work has mostly been built on proprietary models hidden behind opaque API endpoints.
This approach yields experimental results that are not reproducible and non-deterministic, threatening the veracity of outcomes that build on such shaky foundations.
To address this significant shortcoming, we present \rv, the first fully open-source LLM capable of performing high-quality listwise reranking in a zero-shot setting.
Experimental results on the TREC 2019 and 2020 Deep Learning Tracks show that we can achieve effectiveness comparable to zero-shot reranking with \gptthreefive with a much smaller 7B parameter model, although our effectiveness remains slightly behind reranking with \gptfour.
We hope our work provides the foundation for future research on reranking with modern LLMs.
All the code necessary to reproduce our results is available at \url{https://github.com/castorini/rank_llm}.
\end{abstract}

\section{Introduction}

The widespread availability of instruction fine-tuned large language models (LLMs) has led to an explosion of applications in various natural language processing and information retrieval tasks.
In the context of text retrieval, we have seen multiple efforts focused on zero-shot listwise reranking using LLMs~\cite{RankGPT, LRL}, but unfortunately, to date, they have all relied on proprietary models.
While such models support rapid prototyping, particularly when exposed as API endpoints, the reproducibility of experimental results that build on them is suspect---both from the normative perspective of what is ``good science'' and the practical perspective of obtaining reliable and deterministic measurements of experimental results.
It would, of course, be desirable for the community to have access to a fully open-source LLM and associated code infrastructure capable of performing high-quality reranking.

\rv provides exactly this:\
To our knowledge, we present the first open-source large language model for zero-shot listwise document reranking.
Experimental validation on test collections from the TREC 2019 and 2020 Deep Learning Tracks~\cite{dl19,dl20} shows that the effectiveness of our model is on par with zero-shot reranking using \gptthreefive, but slightly worse than reranking with \gptfour.
However, we can achieve these results with a much smaller model with only 7B parameters while still constrained to a \gptthreefive teacher.
We share our model checkpoints and associated code, providing a valuable resource for the research community.

During the process of building \rv, we have gained several important insights that we share: 
First, we confirm that proprietary LLMs are indeed effective at reranking in a zero-shot manner~\cite{RankGPT, LRL}, although they exhibit several shortcomings.
Beyond the obvious issue of non-reproducibility, results from these models are also non-deterministic, which makes them unreliable for rigorous scientific research.
Additionally, proprietary LLMs occasionally fail to follow the requested format in their responses.
In contrast, \rv is open-source, deterministic, and always generates well-formed responses.

Second, we examine the impact of first-stage retrieval methods on downstream reranking effectiveness and find that \rv consistently improves over the baseline retrieved results.
We also find that with an effective first-stage retriever, even a single pass with reranking only the top 20 candidates brings an improvement similar to reranking the top 100 candidates.

Finally, our experiments shed some light on the importance of training strategies that involve data augmentation to ensure model robustness against shuffled candidates or variations in initial retrieval quality.
However, we note that data augmentation techniques affect the quality of model outputs under ``ideal'' conditions, and thus we face an effectiveness--robustness tradeoff.

Our work lays a solid foundation for future research.
By making our models and infrastructure available to the public, we hope to stimulate further exploration and innovation in reranking.
We anticipate that our findings will guide researchers in developing more effective and efficient reranking models.
As the demand for accurate and reliable information retrieval systems continues to grow in this age of retrieval-augmented LLMs, we expect our work to contribute to future advances.

\section{Background and Related Work}

Given a corpus $\mathcal{C}=\{D_1, D_2, ..., D_n\}$ containing a collection of documents and an information need expressed as a query $q$, the task of a \textit{retriever} is to efficiently return a list of $k$ documents from $\mathcal{C}$ that are most relevant to the query $q$ according to some metric such as nDCG or average precision, where $k \ll |\mathcal{C}|$.
The task of a \textit{reranker} is to further improve the quality of the ranked list produced by the retriever or another upstream reranker, according to either the same or a different metric.

Retrievers and rerankers together form multi-stage ranking pipelines for text ranking, which have been studied in the context of transformer models~\cite{duobert, lce} but date back well over a decade \cite{Matveeva_etal_SIGIR2006, Cambazoglu_etal_WSDM2010, Wang_etal_SIGIR2011}.
\citet{monobert} were the first to demonstrate the use of (encoder-only) transformer models for reranking (using BERT) with a simple cross-encoder architecture they called monoBERT.
While neural rerankers had been explored extensively by researchers prior to the advent of BERT, the monoBERT model represented a significant advance in effectiveness; see \citet{Lin_etal_2021_ptr4tr} for a historical overview.

Following monoBERT, other researchers have explored reranking using decoder-only transformer models~\cite{nogueira-dos-santos-etal-2020-beyond} and full encoder--decoder models~\cite{monoT5,rankt5}.
These approaches are effective but require copious amounts of training data in the form of (query, relevant passage) pairs; often, the MS MARCO dataset~\citep{msmarco} is used for such purposes.
Most of the early work on reranking with transformers can be characterized as a {\it pointwise} approach, where the relevance of a particular candidate document is estimated independently of others.

More recently, however, researchers have addressed this shortcoming by incorporating {\it pairwise} and {\it listwise} losses in their cross-encoder approaches~\cite{lce, SqueezeMBERT, rankt5}. 
Using hard negatives in combination with such losses yields systems that are better at reranking in high-precision settings and that align more closely to the first-stage retriever.

In contrast, our work focuses on the zero-shot setting, where the model is not provided any task-specific supervised training (e.g., relevant query--passage pairs).
We build on a recent thread of work~\cite{RankGPT, LRL, PRP} that directly uses LLMs as rerankers in a multi-stage ranking pipeline, primarily focusing on prompt engineering to accomplish the reranking task.
We coin the term ``prompt-decoders'' (in contrast to BERT-style cross-encoders) to characterize this class of rerankers.
Furthermore, since these models are not fine-tuned or benefit from in-context learning, we might describe this type of reranking model as a {\it zero-shot} prompt-decoder.
To use an open-source LLM as a prompt-decoder, \citet{PRP} adopted a pairwise approach since FLAN-UL2 is not capable of reordering a list of input documents.
We find the same shortcoming to be also true for Vicuna, but we address this by using \rgthreefive as its teacher.

Rerankers depend on an upstream source to supply candidate documents, which can be a first-stage retriever or another reranker.
In all our experiments, we rely on a first-stage retriever to generate a candidate list of documents from the corpus.
Researchers have explored a variety of sparse, dense, and hybrid retrieval techniques, but these are not the focus of our study.
We refer interested readers to~\citet{Lin_arXiv2021_repir} and~\citet{Lin_etal_2021_ptr4tr} for an overview of such models.

In another relevant thread, recent work such as InPars~\citep{inpars, inpars-light} and Promptagator~\citep{promptagator} explored using LLMs to generate synthetic queries for documents to craft relevant query--document pairs as training data for retrievers or rerankers.
Similarly, HyDE~\citep{hyde} used LLMs to augment queries by generating hypothetical documents for unsupervised dense retrieval.
Related,~\citet{QuestionAllYouNeedDPR} proposed ART, a novel approach to training a dense passage retriever starting only with questions, which outperforms the standard reference dense retrieval model DPR~\cite{dpr}.
In the emerging paradigm of generative retrieval, \citet{DSIScaling} explored different document representation strategies and found synthetic queries to be necessary for effectiveness as the corpus size increases.
However, all these approaches take advantage of large language models {\it indirectly.}

Finally, we note that rerankers have gained additional prominence in recent months with the introduction of commercially available API endpoints.
Examples include Cohere's Rerank API\footnote{\url{https://cohere.com/rerank}} and Microsoft's Semantic Search API in Azure Cognitive Search.\footnote{\url{https://learn.microsoft.com/en-us/azure/search/semantic-search-overview}}
The existence of these production services suggests that reranking models have attained maturity beyond explorations in research laboratories, and that rerankers address a real-world problem.

\section{Methods}

\subsection{Prompt Design}
\label{sec:prompt}

Recent work~\citep{LRL} has shown that zero-shot \textit{listwise} LLM-based rerankers outperform their \textit{pointwise} counterparts since the former can attend to multiple documents simultaneously to determine their relative positions in a relevance ranking.
We build on this finding and define our ranking problem as follows:\ 
Given a user query $q$ and candidate documents $\{D_1, \ldots, D_n\}$ from the previous stage, the task is to return a reordered list of the input document identifiers that improves a retrieval metric such as nDCG.

Our prompt template for zero-shot listwise reranking is similar to the \rg prompt~\cite{RankGPT}, but accounts for differences between Vicuna and GPT; specifically, we use the default system description for Vicuna.
In addition, we modified the prompt to show that the answer can, and in many cases \emph{should}, deviate from the identity ordering, $[1] > [2] > \ldots > [m]$.
The exact input prompt to Vicuna is shown in \autoref{prompt:1}.

\begin{figure}[t]
\begin{small}
\begin{verbatim}
USER: I will provide you with {num} passages, each
indicated by a numerical identifier []. Rank the
passages based on their relevance to the search
query: {query}.

[1] {passage 1}
[2] {passage 2}
...
[{num}] {passage {num}}

Search Query: {query}.

Rank the {num} passages above based on their 
relevance to the search query. All the passages
should be included and listed using identifiers, in
descending order of relevance. The output format
should be [] > [], e.g., [4] > [2]. Only respond
with the ranking results, do not say any word
or explain.
\end{verbatim}
\end{small}
\caption{User Input for both \rv and our replication of RankGPT.}
\label{prompt:1}
\end{figure}

We prepend the prompt with the system description, which, in Vicuna's case, is ``A chat between a curious user and an artificial intelligence assistant. The assistant gives helpful, detailed, and polite answers to the user's questions.'' 
We hope that aligning our model with the exact prompt setup used to train Vicuna would help generate higher-quality ranked lists for our task.

\subsection{\rv}
\label{sec:rankvicuna}

We leveraged \rgthreefive as a teacher model for Vicuna to prompt-decode high-quality ranked lists.
More specifically, we trained \rv on the ranked lists generated by \rgthreefive for the 100K training set queries provided by~\citet{RankGPT}.
To generate this dataset, the authors randomly sampled 100K queries from the MS MARCO v1 passage ranking training set and retrieved 20 candidates using BM25 for each query using Pyserini~\cite{Pyserini}.
Then, these candidates were passed into \rgthreefive to generate teacher orderings, which we distill down to our student, \rv.
Since both \rgthreefive and \rv are not directly exposed to human-labeled relevant query--passage pairs, our approach can still be considered zero-shot.

To ensure higher quality and more robust trained models, we took the following additional steps:

\begin{itemize}[leftmargin=*]
    \item We did not train on malformed generations. 
    More specifically, examples with incorrect list formatting, missing document identifiers, or repetitions were excluded from the training set.
    This is important as we find that about 12\% of the outputs were malformed, and we desire a model that consistently generates a well-formed ordering.

    \item Besides including the original generations provided by the teacher, which reranks the top 20 results by BM25~\citep{bm25}, we also include a condition where the input order is shuffled.
    Our hope is that this exposes the model to a more complex reordering task while not incurring additional data generation costs.
    However, we still retain the original BM25 input ordering, as we believe it is important to model ``success'', given it is the closest to what the model sees during inference.
    All \rv settings in the rest of the paper involve this data augmentation (DA) process unless specified.
\end{itemize}

\noindent We trained our 7B parameter \rv for two epochs with an effective batch size of 128 and a learning rate of $2 \times 10^{-5}$ in bfloat16.
Training took roughly 80 hours on four NVIDIA RTX A6000 GPUs.
The Vicuna model that served as our initial weights can be found under \texttt{lmsys/vicuna-7b-v1.5} in the HuggingFace Hub.
This model is instruction fine-tuned from Meta's LLaMA-v2 model~\cite{LLaMAv2}.

It is worth noting that the ``out-of-the-box'' Vicuna model, which was {\it not} trained on the \rgthreefive data, completely fails at the reranking task, often simply returning an identity ordering or a malformed generation.

\begin{table*}[t]
\centering \scalebox{0.75}{
\begin{tabular}{lllllll}
\toprule
\toprule
\multicolumn{1}{l}{\multirow{2}{*}{\hspace{175pt}}} & 
\multicolumn{2}{l}
{\multirow{2}{*}
{\begin{tabular}
[c]{@{}c@{}}\textbf{Source} \\
\textbf{Prev.} \hspace{35pt}  \textbf{Top-k}
\end{tabular}}} &
\multicolumn{2}{l}
{\multirow{2}{*}
{\begin{tabular}
[c]{@{}c@{}}\textbf{DL19} \\
\textbf{nDCG@10} \hspace{20pt}  \textbf{MAP@100}
\end{tabular}}} &
\multicolumn{2}{l}
{\multirow{2}{*}
{\begin{tabular}
[c]{@{}c@{}}\textbf{DL20} \\
\textbf{nDCG@10} \hspace{20pt}  \textbf{MAP@100}
\end{tabular}}} \\
\multicolumn{1}{l}{}\ & \multicolumn{1}{l}{}\ & \multicolumn{2}{c}{} & \multicolumn{2}{c}{} \\
\midrule
(1) BM25                                        &
\multicolumn{2}{l}{None}{$|C|$}                 &
0.5058 & 0.2476                                 &
0.4796 & 0.2685                                 \\
(2) Contriever                                  &
\multicolumn{2}{l}{None}{$|C|$}                 &
0.6164 & 0.3163                                 &
0.5986 & 0.3309                                 \\
\midrule
(3) \LRL (\gptthreefive)                &
\multicolumn{2}{l}{BM25}{100}                   &
0.6451$\pm$0.003 & 0.3035$\pm$0.004             &
0.6099$\pm$0.004 & 0.3496$\pm$0.004             \\
\midrule
(4) \rgthreefive                &
\multicolumn{2}{l}{BM25}{100}                   &
0.6855$\pm$0.006 & 0.3335$\pm$0.002             &
0.6202$\pm$0.005 & 0.3525$\pm$0.002             \\
(5) \rgfour                  &
\multicolumn{2}{l}{BM25}{100}                   &
0.7500$\pm$0.002 & 0.3703$\pm$0.004             &
0.7036$\pm$0.004 & 0.4134$\pm$0.004             \\
\midrule
(6) PRP-Sliding-10 (FLAN-T5-XXL)                &
\multicolumn{2}{l}{BM25}{100}                   &
0.6700 &\textbf{-}                              &
0.6735 &\textbf{-}                              \\
(7) PRP-Sliding-10 (FLAN-UL2)                   &
\multicolumn{2}{l}{BM25}{100}                   &
0.7265 & \textbf{-}                             &
0.7046 & \textbf{-}                             \\
\midrule
(8) PRP-Sliding-10 (Vicuna)                     &
\multicolumn{2}{l}{BM25}{100}                   &
0.5606   & 0.2735                               &
0.5367   & 0.2990                               \\
\midrule
(9) \rv                                         &
\multicolumn{2}{l}{BM25}{100}                   &
0.6682 & 0.3316                                 &
0.6549 & 0.3789                                 \\
(10) \rv & 
\multicolumn{2}{l}{SPLADE++ ED}{100}            & 
0.7459 & 0.4416                                 &
0.7473 & 0.5183                                 \\
\bottomrule 
\bottomrule                                            
\end{tabular}}
\caption{
nDCG@10 and MAP@100 on DL19 and DL20 for different reranking pipelines, with BM25 and Contriever as baselines. Each reranker uses the top 100 retrieved results of the previous stage as input. Rows (3--4) and row (5) represent averages of six and three runs, respectively.
We directly copied results in rows (6--7) from \citet{PRP}.
All other results are from our own experiments.}
\label{tab:1}
\end{table*}

\section{Experimental Setup}

To demonstrate the effectiveness of \rv, we compared it with existing representative unsupervised ranking methods (BM25 and Contriever) as well as our replications of two closed-source prompt-decoder models: LRL~\cite{LRL} with \gptthreefive and \rg~\cite{RankGPT}, with both \gptthreefive and \gptfour, which we refer to as \rgthreefive and \rgfour, respectively.
\gptthreefive refers to the model dubbed \texttt{gpt-3.5-turbo} in the Open\-AI suite while \gptfour refers to \texttt{gpt-4}.
We also compared \rv with our replication of PRP-Sliding-10 from~\citet{PRP}, albeit with Vicuna (7B parameters).
For these experiments, we used Vicuna instead of FLAN-T5 or FLAN-UL2 because we wanted an apples-to-apples comparison with the same base LLM.
Additionally, we note that the FLAN mixture, used to \emph{pretrain} the models, includes the MS MARCO QA task,\footnote{\url{https://github.com/google-research/FLAN/blob/e9e4ec6e2701182c7a91af176f705310da541277/flan/v2/flan\_collection\_info.csv\#L1032}} thereby rendering the results suspect from the perspective of zero-shot retrieval.

We evaluated our methods using test collections from the TREC 2019 and 2020 Deep Learning Tracks~\cite{dl19, dl20}, using query and relevance judgments from the passage retrieval tasks.
These tasks use the MS MARCO v1 passage corpus~\citep{msmarco}, which contains 8.8 million passages.
For convenience, we refer to these datasets as DL19 and DL20.
We report effectiveness in terms of nDCG@10 and average precision at a rank cutoff of 100 (denoted MAP@100).

The context size is 4096 for Vicuna and \gptthreefive and 8192 for GPT\textsubscript{4}.
To reorder the top 100 candidates for each query given these context sizes, we used a sliding window similar to \rg and \LRL. 
In our experiments, we have adopted the same values as \rg (window size 20, stride 10) to isolate the impact of window and stride size in our comparisons.

Unlike \rv, we (surprisingly) observe non-deterministic outputs for \gptthreefive and \gptfour, even with a temperature of zero.
For these two models, we report the mean over six and three runs, respectively, with $99\%$ confidence intervals.
We limited the number of \gptfour runs to three due to our computation budget.

In all our reranking experiments, we replaced any reference of the form $[n]$ in the passages with $(n)$ to avoid confusing the models.
We also leveraged \texttt{ftfy}'s \texttt{fix\_text} method to preprocess any input sent to the rerankers.

\begin{table*}[]
\centering \scalebox{0.75}{
\begin{tabular}{lccccc}
\toprule
\toprule
& \textbf{OK} & \textbf{Wrong Format} & \textbf{Repetition} & \textbf{Missing} & \textbf{Total} \\
\midrule 
\rgthreefive& 838.67 & 0     & 1.16 & 33.16 & 873   \\ 
\rgfour  & 830.33 & 40.67 & 1.67 & 0.33  & 873   \\
\rv                         & 873    & 0     & 0    & 0     & 873   \\
\bottomrule
\bottomrule
\end{tabular}}
\caption{The number of malformed responses for each reranking method.
Reported numbers for \rgthreefive and \rgfour are averages of three and six runs, respectively.}
\label{tab:correctness}
\end{table*}

\section{Results}

\autoref{tab:1} compares different reranking pipelines using data from DL19 and DL20.
Rows~(1) and (2) report baselines using two first-stage retrievers, BM25 and Contriever~\cite{contriever}.
The remaining rows (besides the last one) report the results of using zero-shot LLM rerankers to reorder top 100 candidate documents retrieved by BM25.
Rows~(6) and (7) show scores of two variants of PRP-Sliding-10, FLAN-T5-XXL and FLAN-UL2, directly copied from~\citet{PRP}.
The final row represents our best system, where we apply \rv to rerank the top 100 candidates generated by SPLADE++ EnsembleDistil~\cite{splade}, a state-of-the-art neural first-stage sparse retrieval method.

As expected, all LLM rerankers outperform the baseline (first-stage) methods.
The effectiveness of \rv, with 7B parameters, is on par with the effectiveness of \rgthreefive, with 175B parameters.
Specifically, compared to its teacher \rgthreefive, \rv achieves higher scores on DL20 but slightly lower scores on DL19.
Compared with another zero-shot reranking method, LRL, which uses \rgthreefive, \rv demonstrates considerably higher effectiveness on both DL19 and DL20.

We note that PRP-Sliding-10 (FLAN-T5-XXL) with 11B parameters is comparable to \rv both in terms of model size and effectiveness. 
Other than being fully open-source, our main advantage over PRP-Sliding-10 (FLAN-T5-XXL) is the prompt cost:\ to bring the top 10 most relevant candidates to the top of the list, PRP-Sliding-10 (FLAN-T5-XXL) requires each passage to be included in $\sim$40 prompts on average.
In contrast, we only require two prompts for our listwise approach with a sliding window of size 20 and a stride of 10.
Furthermore, training on the FLAN mixture, which includes the MS MARCO QA task, calls into question the validity of PRP-Sliding-10 (FLAN-T5-XXL) as a true zero-shot method.
We suspect this to be a contributing factor to the effectiveness gap between PRP-Sliding-10 (FLAN-T5-XXL) and PRP-Sliding-10 (Vicuna).

Not surprisingly, both \rgfour (rumored to contain more than 1T parameters) and PRP-Sliding-10 (FLAN-T5-UL2) with 20B parameters outperform \rv.
This could be because, in addition to the differences in model sizes, the effectiveness of \rv is bounded by its teacher, \rgthreefive.

Finally, in row (10), we used \rv to rerank the top 100 candidates from SPLADE++ EnsembleDistil instead of BM25.
This combination achieves effectiveness on par with \rgfour with an open-source model that is more than two orders of magnitude smaller.

\begin{table*}[t]
\centering \scalebox{0.75}{
\begin{tabular}{lllllll}
\toprule
\toprule
\multicolumn{1}{l}{\multirow{2}{*}{\hspace{140pt}}} & 
\multicolumn{2}{l}
{\multirow{2}{*}
{\begin{tabular}
[c]{@{}c@{}}\textbf{Source} \\
\textbf{Prev.} \hspace{70pt}  \textbf{Top-k}
\end{tabular}}} &
\multicolumn{2}{l}
{\multirow{2}{*}
{\begin{tabular}
[c]{@{}c@{}}\textbf{DL19} \\
\textbf{nDCG@10} \hspace{30pt}  \textbf{MAP@100}
\end{tabular}}} &
\multicolumn{2}{l}
{\multirow{2}{*}
{\begin{tabular}
[c]{@{}c@{}}\textbf{DL20} \\
\textbf{nDCG@10} \hspace{30pt}  \textbf{MAP@100}
\end{tabular}}} \\
\multicolumn{1}{l}{}\ & \multicolumn{1}{l}{}\ & \multicolumn{2}{c}{} & \multicolumn{2}{c}{} \\
\midrule
(1a) BM25                                       &
\multicolumn{2}{l}{None}{$|C|$}                 &
0.5058 & \hspace{42pt} 0.2476                   &
0.4796 & \hspace{42pt} 0.2685                   \\
(1b) \rv                                        &
\multicolumn{2}{l}{BM25}{20}                    &
0.6164 & \hspace{42pt} 0.2867                   &
0.5986 & \hspace{42pt} 0.3194                   \\
(1c) \rv                                        &
\multicolumn{2}{l}{BM25}{100}                   &
0.6682 & \hspace{42pt} 0.3316                   &
0.6549 & \hspace{42pt} 0.3789                   \\
\midrule
(2a) BM25 + RM3                                 & 
\multicolumn{2}{l}{None}{$|C|$}                 &
0.5216 & \hspace{42pt} 0.2807                   &
0.4896 & \hspace{42pt} 0.2821                   \\
(2b) \rv                                        & 
\multicolumn{2}{l}{BM25 + RM3}{20}              &
0.6053 & \hspace{42pt}  0.3110                  &
0.5825 & \hspace{42pt}  0.3323                  \\
(2c) \rv                                        & 
\multicolumn{2}{l}{BM25 + RM3}{100}             &
0.6588 & \hspace{42pt} 0.3573                   &
0.6567 & \hspace{42pt} 0.3991                   \\
\midrule
(3a) OpenAI ada2                                &
\multicolumn{2}{l}{None}{$|C|$}                 &
0.7035 & \hspace{42pt} 0.4151                   &
0.6759 & \hspace{42pt} 0.4587                   \\
(3b) \rv                                        &
\multicolumn{2}{l}{OpenAI ada2}{20}             &
0.7448 & \hspace{42pt} 0.4398                   &
0.7101 & \hspace{42pt} 0.4718                   \\
(3c) \rv                                        &
\multicolumn{2}{l}{OpenAI ada2}{100}            &
0.7374 & \hspace{42pt} 0.4409                   &
0.7210 & \hspace{42pt} 0.4755                   \\
\midrule
(4a) DistillBERT KD TASB                        &
\multicolumn{2}{l}{None}{$|C|$}                 &
0.7210 & \hspace{42pt} 0.4050                   &
0.6854 & \hspace{42pt} 0.4520                   \\
(4b) \rv                                        &
\multicolumn{2}{l}{DistillBERT KD TASB}{20}     &
0.7588 & \hspace{42pt} 0.4121                   &
0.7404 & \hspace{42pt} 0.4648                   \\
(4c) \rv                                        &
\multicolumn{2}{l}{DistillBERT KD TASB}{100}    &
0.7551 & \hspace{42pt} 0.4170                   &
0.7049 & \hspace{42pt} 0.4620                   \\
\midrule
(5a) SPLADE++ ED                                &
\multicolumn{2}{l}{None}{$|C|$}                 &
0.7308 & \hspace{42pt} 0.4464                   &
0.7197 & \hspace{42pt} 0.4826                   \\
(5b) \rv                                        &
\multicolumn{2}{l}{SPLADE++ ED}{20}             &
0.7532 & \hspace{42pt} 0.4491                   &
0.7455 & \hspace{42pt} 0.5150                   \\
(5c) \rv                                        &
\multicolumn{2}{l}{SPLADE++ ED}{100}            &
0.7459 & \hspace{42pt} 0.4416                   &
0.7473 & \hspace{42pt} 0.5183                   \\
\bottomrule 
\bottomrule                                                           
\end{tabular}}
\caption{
nDCG@10 and MAP@100 for \rv with different first-stage candidate generation methods. For each method, reranking is performed using the top 20 or 100 candidates.}
\label{tab:first_stages}
\end{table*}

\autoref{tab:correctness} shows the number of malformed responses generated by the \rg variants and \rv, which we have grouped into the following categories:

\begin{enumerate}[leftmargin=*]

\item \textit{Wrong Format}: includes responses that do not follow the requested format.
For example, when \rgfour refuses to generate a sorted list, its response falls into this category.

\item \textit{Repetition}: includes responses that contain repeated document ids.

\item \textit{Missing}: includes responses with missing document ids.

\end{enumerate}

\noindent Since \rv is deterministic, we report the results of a single run.
For every request in this run, \rv returned a correctly formatted response.
In contrast, for \rgthreefive and \rgfour, we averaged the results of six and three runs, respectively.
Both \rg methods occasionally return malformed responses. 
Most of the malformed responses from \rgthreefive are missing documents in the ordered list; when malformed, \rgfour mostly refuses to rank.
Repetition is a rare problem for both \rg methods.

\section{Ablation Studies} 

\subsection{First-Stage Candidate Generation}

To evaluate the impact of the quality and quantity of the generated candidates on the final results, we repeated our experiments with the following five first-stage retrieval methods using either top 20 or top 100 retrieved results: (1) BM25~\cite{bm25}, (2) BM25+RM3~\cite{RM3}, (3) OpenAI ada2~\cite{OpenAIEmbed, LuceneIsAll}, (4) DistillBERT KD TASB~\cite{tasb}, (5)  SPLADE++ EnsembleDistil (ED)~\cite{spladepp}.
The first two represent strong traditional ``bag-of-words'' retrieval baselines; the others represent a sample of effective neural first-stage retrievers that are commonly seen in research studies today.
OpenAI ada2 and DistillBERT KD TASB are dense retrieval methods, while SPLADE++ ED is a sparse one.

Our experiment shows that as the first-stage effectiveness increases, additional improvements from \rv decrease (see \autoref{tab:first_stages}).
For example, while \rv over the top 100 BM25 candidates improves effectiveness by 30\%--45\% for all metrics, the improvement for SPLADE++ ED is only 2\%--4\% for the same metrics.
This is a commonly noted phenomenon across multi-stage ranking systems~\cite{EMD, SqueezeMBERT, CT22}.

Comparing top 20 vs.\ top 100 results shows that reranking more candidates generally results in a higher MAP@100.
However, in cases where the first-stage effectiveness is ``good enough'', rows (3--5) for DL19 and rows (4--5) for DL20, reranking only the top 20 candidates achieves an nDCG@10 score on par with reranking the top 100 candidates.

\begin{table*}
\centering \scalebox{0.75}{
\begin{tabular}{lllllll}
\toprule
\toprule
\multicolumn{1}{l}{\multirow{2}{*}{\hspace{140pt}}} & 
\multicolumn{2}{l}
{\multirow{2}{*}
{\begin{tabular}
[c]{@{}c@{}}\textbf{Source} \\
\textbf{Prev.} \hspace{70pt}  \textbf{Top-k}
\end{tabular}}} &
\multicolumn{2}{l}
{\multirow{2}{*}
{\begin{tabular}
[c]{@{}c@{}}\textbf{DL19} \\
\textbf{nDCG@10} \hspace{20pt}  \textbf{MAP@100}
\end{tabular}}} &
\multicolumn{2}{l}
{\multirow{2}{*}
{\begin{tabular}
[c]{@{}c@{}}\textbf{DL20} \\
\textbf{nDCG@10} \hspace{20pt}  \textbf{MAP@100}
\end{tabular}}} \\
\multicolumn{1}{l}{} & \multicolumn{2}{l}{} & \multicolumn{2}{l}{} & \multicolumn{2}{l}{} \\
\midrule
(1a) \rv                                        &
\multicolumn{2}{l}{BM25}{100}                   &
0.6682 & 0.3316                                 &
0.6549 & 0.3789                                 \\
(1b) \rv                                        &
\multicolumn{2}{l}{Shuf.\ BM25}{100}            &
0.6702$\pm$0.009 & 0.2977$\pm$0.006             &
0.6537$\pm$0.006 & 0.3553$\pm$0.006             \\
(1c) \rv                                        &
\multicolumn{2}{l}{SPLADE++ ED}{100}            &
0.7459 & 0.4416                                 &
0.7473 & 0.5183                                 \\
(1d) \rv                                        &
\multicolumn{2}{l}{Shuf.\ SPLADE++ ED}{100}     &
0.7271$\pm$0.009 & 0.3860$\pm$0.008             &
0.7071$\pm$0.007 & 0.4312$\pm$0.006             \\
\midrule
(2a) \rv (w/o DA)                               &
\multicolumn{2}{l}{BM25}{100}                   &
0.6612 & 0.3254                                 &
0.6420 & 0.3612                                 \\
(2b) \rv (w/o DA)                               &
\multicolumn{2}{l}{Shuf.\ BM25}{100}            &
0.5893$\pm$0.017 & 0.2666$\pm$0.011             &
0.5293$\pm$0.010 & 0.2754$\pm$0.007             \\
(2c) \rv (w/o DA)                               &
\multicolumn{2}{l}{SPLADE++ ED}{100}            &
0.7653 & 0.4672                                 &
0.7536 & 0.5180                                 \\
(2d) \rv (w/o DA)                               &
\multicolumn{2}{l}{Shuf.\ SPLADE++ ED}{100}     &
0.5893$\pm$0.010 & 0.3289$\pm$0.009             &
0.5373$\pm$0.020 & 0.3406$\pm$0.013             \\
\bottomrule
\bottomrule
\end{tabular}}
\caption{
nDCG@10 and MAP@100 of two variants of \rv with different first-stage candidate generation methods.
For each method, reranking is performed using top 100 candidates from the previous step on six shuffled orderings.
We report average metrics and with 99\% confidence intervals.}
\label{tab:shuffling}
\end{table*}
\pgfplotsset{compat=1.17}
\begin{figure*}
    \centering
    \begin{tikzpicture}
        \begin{axis}[
            xlabel={Number of Sliding Window Passes},
            ylabel={nDCG@10},
            xtick={0,1,2,3,4,5,6,7,8,9,10},
            xticklabels={0,1,2,3,4,5,6,7,8,9,10},
            width=12cm,
            height=5cm,
            legend style={
                at={(1.02,0.5)},
                anchor=west,
                legend columns=1,
                /tikz/every even column/.append style={column sep=5pt},
            },
        ]

        \addplot[blue,mark=square*] coordinates {
            (0, 0.5058)
            (1, 0.6682)
            (2, 0.6837)
            (3, 0.6818)
            (4, 0.6797)
            (5, 0.6817)
            (6, 0.6808)
            (7, 0.6804)
            (8, 0.6795)
            (9, 0.6804)
            (10, 0.6795)
        };
        \addlegendentry{RankVicuna on \emph{DL19}}

        \addplot[red,mark=square*] coordinates {
            (0, 0.5058)
            (1, 0.5305)
            (2, 0.5447)
            (3, 0.5549)
            (4, 0.5585)
            (5, 0.5594)
            (6, 0.5602)
            (7, 0.5597)
            (8, 0.5604)
            (9, 0.5616)
            (10, 0.5606)
        };
        \addlegendentry{PRPVicuna on \emph{DL19}}

        \addplot[blue,mark=*,dashed] coordinates {
            (0, 0.4796)
            (1, 0.6549)
            (2, 0.6572)
            (3, 0.6581)
            (4, 0.6604)
            (5, 0.6592)
            (6, 0.6590)
            (7, 0.6590)
            (8, 0.6590)
            (9, 0.6590)
            (10, 0.6590)
        };
         \addlegendentry{RankVicuna on \emph{DL20}}

        \addplot[red,mark=*,dashed] coordinates {
            (0, 0.4796)
            (1, 0.5106)
            (2, 0.5214)
            (3, 0.5277)
            (4, 0.5329)
            (5, 0.5332)
            (6, 0.5332)
            (7, 0.5339)
            (8, 0.5350)
            (9, 0.5361)
            (10, 0.5367)
        };
        \addlegendentry{PRPVicuna on \emph{DL20}}
        \end{axis}
    \end{tikzpicture}
    \caption{Comparing the effectiveness of RankVicuna vs.\ PRPVicuna on DL19 and DL20, varying the number of times the ranked list is progressively refined. The zeroth pass corresponds to the BM25 run.}
    \label{fig:step-graph}
\end{figure*}
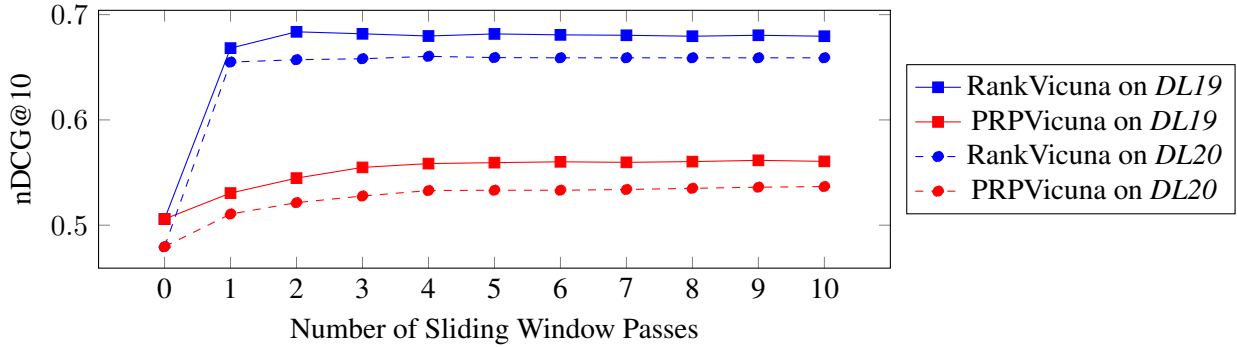

\subsection{Data Augmentation}

Section~\ref{sec:rankvicuna} discussed the training process of \rv, highlighting the use of data augmentation (DA) as a crucial step in our training pipeline.
To recap, the DA process involves shuffling the input order of the documents and permuting the original generations provided by the teacher.
This step exposes the model to a more complex reordering task, which hopefully enhances its robustness and effectiveness. 

In this section, we study the dependence of \rv on the order of generated candidates.
We compared two versions of the model:\ (1) the default version trained using Data Augmentation (DA), and (2) a variant trained without DA.
Experimental results are shown in \autoref{tab:shuffling}.

Using BM25 as the first stage, our experiments show that \rv without DA results in worse effectiveness than using \rv with DA.
When we replace BM25 with SPLADE++ ED, \rv without DA outperforms \rv with DA.
While data augmentation can cause a small drop in effectiveness (depending on the first stage), it makes the model less vulnerable to poor quality candidates (whether intentional or not), as shown by~\citet{PRP} in methods like PRP-Sliding-10 and \rgthreefive.

To showcase this vulnerability, we provided both model variants with shuffled candidate documents (rows \textit{b} and \textit{d}).
The results show that the model without DA exhibited a significant effectiveness drop (up to 34\%) and higher variance among different runs.
In contrast, the default model, which is more robust due to its exposure to a more complex reordering task, better retained its effectiveness (comparing rows \textit{b} vs.\ \textit{a} and \textit{d} vs.\ \textit{c}, respectively, for each version).

\subsection{Effect of Progressive Reranking}

Finally, \autoref{fig:step-graph} compares the effectiveness of two reranking methods, \rv and a variant of PRP-Sliding from~\citet{PRP}, we call PRPVicuna, on two datasets, DL19 and DL20.
The $x$-axis represents the number of sliding window passes, ranging from 0 to 10, and the $y$-axis represents the nDCG@10 score.
We plot four curves, each representing a combination of a reranking method and a dataset. 
The solid lines show results on DL19 and the dashed lines show results on DL20.
The blue lines represent the \rv method and the red lines represent the PRPVicuna method~\cite{PRP}.

We see that, for both datasets, \rv consistently outperforms PRPVicuna. 
The nDCG@10 score for \rv on DL19 starts at 0.5058 and increases to 0.6837 at the second pass, remaining relatively stable thereafter.
The score for \rv on DL20 follows a similar pattern, starting at 0.4796 and rising to about 0.6604 at pass four, albeit at a slower pace after the first pass.
On the other hand, the nDCG@10 scores for PRPVicuna on both datasets increase gradually with each pass but remain far below \rv.

This plot suggests that \rv is more effective than PRPVicuna and that multiple passes of the sliding window have a minimal impact as an effectiveness boost for \rv.
It is also worth noting that a single pass of reranking with both methods takes about the same time, around 30 seconds per query using a batch size of one on an RTX A6000 GPU.
These results show that \rv is much more efficient and achieves quicker convergence to the best possible results.
This is likely because PRPVicuna handles only two passages at a time, whereas \rv attends to 20 passages simultaneously, resulting in more effective relevance estimation.

\section{Conclusion}

In this study, we introduce \rv, a listwise zero-shot reranking approach powered by an open-source large language model, Vicuna.
Experimental studies show that our model achieves effectiveness on par with much larger models.
We also quantitatively demonstrated the stability of \rv results compared to closed-source counterparts.

Along the way, we explored many aspects of prompt-decoder models for reranking, including the impact of first-stage retrievers on downstream effectiveness.
Our work also sheds light on the importance of data augmentation for system robustness, which plays a vital role in ensuring stability in the face of document shuffling and variations in initial retrieval quality.

In summary, \rv advances zero-shot reranking for information retrieval, demonstrating the potential of large language models to enhance search effectiveness, even in data-scarce settings.
We are able to achieve high-quality reranking using fully open-source models, which provides a firm foundation for the rest of the research community to build on.
As we further refine and expand these techniques, we anticipate exciting opportunities for integrating large language models into end-to-end information access applications.

\section*{Acknowledgments}

This research was supported in part by the Natural Sciences and Engineering Research Council (NSERC) of Canada.

\bibliography{custom}
\bibliographystyle{acl_natbib}
\end{document}